\input phyzzx
\vsize=23.5cm
\hsize=16.5cm
\voffset=0cm
\hoffset=0cm
\singlespace
\sequentialequations

\fourteenpoint

\centerline{\bf
QCD Phase Transition at high Temperature in Cosmology 
}

\twelvepoint
\twelverm
\tenpoint
\tenrm

\vskip8pt

\centerline{
H.~Suganuma$^a$, H.~Ichie$^a$, H.~Monden$^a$, S.~Sasaki$^a$, 
M.~Orito$^b$, T.~Yamamoto$^b$ and T.~Kajino$^c$
}
\vskip4pt

\tenpoint
\tenrm

\centerline{\it 
a) Research Center for Nuclear Physics (RCNP), Osaka University, 
Osaka 567, Japan}

\centerline{\it 
b) Dept.~of~Astronomical~Science, 
The~Graduate~Univ.~for~Advanced~Studies, 
Tokyo 181, Japan}

\centerline{\it 
c) Division of Theoretical Astrophysics, 
National Astronomical Observatory (NAO), and} 
\baselineskip=12pt
\centerline{\it 
The~Graduate~Univ.~for~Advanced~Studies; 
Dept.~of~Astronomy, Tokyo~Univ., Tokyo 181, Japan}

\vskip5pt
\baselineskip=8pt

The cosmological QCD phase transition is studied in terms of the color 
confinement at finite temperature using the dual Higgs theory of QCD. 
The confinement force is largely reduced 
at high temperature, which leads to the swelling of hadrons. 
We derive analytical formulae for the surface tension and the 
boundary thickness of the mixed phase 
from the effective potential at $T_c$.
We predict a large reduction of the glueball mass near $T_c$. 
We investigate also the process of the hadron-bubble formation 
in the early Universe.

\vskip5pt


\noindent
{\bf 1. QCD, Color Confinement and Dual Higgs Mechanism}

Quantum chromodynamics (QCD) is the fundamental theory 
of the strong interaction. 
In spite of the simple form of the QCD lagrangian, 
it miraculously provides quite various phenomena 
like color confinement, dynamical chiral-symmetry breaking (D$\chi$SB), 
non-trivial topologies, quantum anomalies and so on.
Therefore, QCD can be regarded as 
{\it an interesting miniature of the history of the Universe}, 
where a quite simple Big Bang also created various things 
including galaxies, stars, lives and thinking reeds. 
This is the most attractive point of the QCD physics.

As a modern progress in QCD, 
confinement physics is providing an important current 
of the hadron physics in '90s, since recent lattice QCD 
studies shed a light on the confinement mechanism. 
In the 't~Hooft abelian gauge [1], 
QCD is reduced into an abelian gauge theory with the QCD-monopole, 
which appears from a hedgehog configuration 
corresponding to the non-trivial homotopy group 
$\pi_2({\rm SU}(N_c)/{\rm U}(1)^{N_c-1})=Z_\infty ^{N_c-1}$ 
on the nonabelian manifold. 
In this gauge, the nonperturbative QCD (NP-QCD) vacuum is 
described as the dual Higgs phase with QCD-monopole condensation.
Due to the dual Meissner effect, color-electric flux is 
squeezed like a one-dimensional flux-tube, 
which leads to the linear quark confinement potential [1].
Thus, the origin of color confinement can be recognized 
as the dual Higgs mechanism by monopole condensation.

Many recent studies based on the lattice QCD [4] 
show QCD-monopole condensation in the confinement phase 
and abelian (monopole) dominance for NP-QCD, e.g., 
linear confinement potential, D$\chi$SB and instantons [4].
Abelian (monopole) dominance means that the essence of NP-QCD 
is described only by abelian (monopole) variables in the abelian 
gauge [4]. 
Therefore the condensed monopole in the 't~Hooft abelian gauge is 
nothing but the {\it relevant collective mode for NP-QCD}, 
and the NP-QCD vacuum can be identified as 
the dual-superconductor in a realistic sense.

As a remarkable fact in the duality physics, 
these are {\it two ``see-saw relations'' between the electric and 
magnetic sectors}. 
\nextline
(1) Due to the Dirac condition $eg=4\pi$ [1] in QCD between 
the electric charge $e$ and the magnetic charge $g$, 
a strong-coupling system in one sector corresponds to 
{\it a weak-coupling system in the other sector.}
\nextline
(2) The long-range confinement system corresponds to 
{\it a short-range interaction system in the other sector.
}
\nextline
As the most attractive point in the dual Higgs theory, 
the highly-nonlocal confinement system can be 
described by {\it a short-range interaction theory with 
the dual variables.} 

\vskip2pt

\noindent
{\bf 2. QCD Phase Transition in Dual Ginzburg-Landau Theory}

The dual Ginzburg-Landau (DGL) theory is the QCD effective theory 
based on the dual Higgs mechanism, and can be derived from 
the QCD lagrangian and the QCD nature 
(monopole condensation and abelian dominance).
The DGL lagrangian [1-3] for the pure-gauge system is described with 
the dual gauge field $B_\mu$ and the QCD-monopole field $\chi$, 
\nextline
\centerline{$
{\cal L}_{\rm DGL}={\rm tr} \left\{
-{1 \over 2}(\partial _\mu  B_\nu -\partial _\nu B_\mu )^2
+[\hat{D}_\mu, \chi]^{\dag}[\hat{D}^\mu, \chi]
-\lambda ( \chi^{\dag} \chi -v^2)^2 \right\},
$}
\nextline
where $\hat{D}_\mu \equiv \hat{\partial}_\mu +i g B_\mu$ 
is the {\it dual covariant derivative.} 

The dual gauge field 
$B_\mu \equiv \vec B_\mu \cdot \vec H = B_\mu^3 T^3+B_\mu^8 T^8$ 
is defined on the {\it dual gauge manifold} 
${\rm U(1)}_m^3 \times {\rm U(1)}_m^8$ [1-3], 
which is the dual space of the maximal torus subgroup
${\rm U(1)}_e^3 \times {\rm U(1)}_e^8$ 
embedded in the original gauge group ${\rm SU(3)}_c$.
The abelian field strength tensor is written as  
$F_{\mu \nu}={}^* (\partial\wedge B)_{\mu\nu}$, 
so that the role of the electric and magnetic fields are interchanged 
in comparison with the ordinary $A_\mu$ description.

The QCD-monopole field $\chi$ is defined as 
$\chi \equiv \sqrt{2} \sum_{\alpha=1}^3 \chi_\alpha E_\alpha$ [3].
In the QCD-monopole condensed vacuum with $|\chi_\alpha| =v$, 
the dual gauge symmetry ${\rm U(1)}_m^3 \times {\rm U(1)}_m^8$ 
is spontaneously broken. 
Through the dual Higgs mechanism, 
the dual gauge field $B_\mu$ acquires its mass $m_B = \sqrt{3}g v $, 
whose inverse provides the hadron-flux-tube radius. 
The dual Meissner effect causes the color-electric field 
excluded from the QCD vacuum, which leads to color confinement. 
The QCD-monopole fluctuations 
$\tilde \chi_\alpha \equiv \chi_\alpha - v$ ($\alpha$=1,2,3) 
also acquire their mass $m_\chi= 2 \sqrt{\lambda} v$ in the 
QCD-monopole condensed vacuum. 
As a relevant prediction, one QCD-monopole fluctuation 
$\tilde \chi_ \equiv \sum_{\alpha=1}^3 \tilde \chi_\alpha$ appears as 
a color-singlet scalar glueball in the confinement phase, 
although the dual gauge field $B_\mu$ and 
the other two combinations of the QCD-monopole fluctuation 
are not color-singlet and cannot be observed.

In the DGL theory, the QCD phase transition is characterized by 
QCD-monopole condensate $\bar \chi \equiv |\chi_\alpha|$,
which is an order parameter on the confinement strength. 
For the study of the QCD phase transition, we formulate 
the effective potential $V_{\rm eff}(\bar \chi;T)$ 
at finite temperature as the function of $\bar \chi$ 
using the {\it quadratic source term} 
to avoid the imaginary scalar-mass problem [1-3].

We study the relation between $V_{\rm eff}(\bar \chi;T_c)$ and 
the surface tension $\sigma$ [1,3], which characterizes the strength of 
the first order in the phase transition, and is very important 
for the shape of the boundary surface in the mixed phase.
Using the {\it sine-Gordon (SG) kink ansatz} [1,3] 
for the boundary profile, 
$
\bar \chi (z)= \bar \chi_c \tan ^{-1}e^{z/\delta}, 
$
we derive simple formulae for the surface tension 
$
\sigma \simeq {4\sqrt{3} \over \pi} \sqrt{h} \bar \chi_c
\simeq (112 {\rm MeV})^3, 
$
and the phase-boundary thickness 
$
2\delta \simeq {2\sqrt{3} \over \pi} \bar \chi_c/\sqrt{h} 
\simeq 3.4 {\rm fm}, 
$
where $h$ is the ``barrier height" between the two minima in 
$V_{\rm eff}(\bar \chi;T_c)$.

At high temperature, the hadron string tension 
$k(T)$ becomes smaller and drops rapidly near $T_c$ [2,3].
Therefore, the slope of the inter-quark potential is reduced 
and the inter-quark distance inside hadrons 
increases at high temperature. 
In addition, the color-electric field spreads according 
to the decrease of $m_B$.
Thus, {\it the reduction of the confinement force leads to 
the swelling of hadrons at high temperature} [3].

We predict also {\it a large mass reduction on 
the QCD-monopole (a scalar glueball) near $T_c$.} 
{\it The violent excitation of QCD-monopoles (scalar glueballs) 
with a reduced mass would promote the QCD phase transition.}

\vskip2pt

\noindent
{\bf 3. Hadron Bubble Formation in the Early Universe}

Finally, we study the hadron bubble formation [3,5] 
in the early Universe using the DGL theory [3].
As Witten pointed out, 
if the QCD phase transition is of the first order, the hadron and QGP 
phases should coexist in the early Universe. 
During the mixed-phase period, there appears 
the inhomogeneity on the baryon density distribution [5], 
which can strongly affects the primordial nucleo-synthesis.

The hadron bubble created in the supercooling QGP phase 
is described by the SG-kink type profile of 
the QCD-monopole condensate [3],
$
\bar \chi (r;R)= C \tan^{-1}e^{(R-r)/\delta}, 
$
where $R$ and $2\delta$ correspond to the hadron-bubble radius 
and the phase-boundary thickness, respectively. 
The total energy of the hadron bubble with radius $R$ can be 
estimated using $V_{\rm eff}(\bar \chi;T)$, 
$
E(R;T) = 4 \pi \int_0^{\infty}drr^2
\{ 3( {d\bar \chi(r;R) \over dr})^2 + V_{\rm eff}(\bar \chi ;T)\}, 
$
which is roughly estimated as the sum of the positive 
surface term and the negative volume term [3].
The hadron-bubble energy $E(R;T)$ takes a maximal value 
at a critical radius $R_c$. Hence, the hadron bubbles 
with $R < R_c$ collapse, and only large hadron bubbles 
with $R > R_c$ grow up with radiating the shock wave [3,5]. 
On the other hand, the creation probability 
of large bubbles is strongly suppressed [3] because of 
the thermodynamical factor 
$P(T) \equiv {\rm exp} \{-{4 \pi \over 3} R_{\rm c}(T)^3 h(T)/T \}$, 
with a large barrier height $h(T)$ in $V_{\rm eff}(\bar \chi;T)$. 
Thus, the only small bubbles are created practically, 
although its radius should be larger than $R_{\rm c}$ [3].

As the temperature decreases, the smaller hadron bubbles 
are created, while the bubble formation rate becomes larger [3].
Then, we can imagine how the QCD phase transition 
happens in the Big Bang scenario: 
(a) Slightly below $T_{\rm c}$, 
only large hadron bubbles appear, but the creation rate is quite small.
(b) As temperature is lowered by the expansion of the Universe, 
smaller bubbles are created with much formation rate.
During this process, the created hadron bubbles expand with 
radiating shock wave, which reheats the QGP phase around them.
(c) Near $T_{\rm low}$, many small hadron bubbles 
are violently created in the unaffected region free from the shock wave. 
(d) The QGP phase pressured by the hadron phase 
is isolated as high-density QGP bubbles [3], 
which provide the baryon density fluctuation [5]. 
\nextline
Thus, the numerical simulation using the DGL theory would tell 
how the hadron bubbles appear and evolve 
quantitatively in the early Universe.

\vskip3pt

\tenpoint
\tenrm
\baselineskip=12pt

{\bf References}

\item{\rm 1.} H.~Suganuma, S.~Sasaki and H.~Toki, 
Nucl.~Phys.~{\bf B435} (1995) 207. 
\nextline
H.~Suganuma, S.~Sasaki, H.~Toki and H.~Ichie, 
Prog.Theor.Phys.(Suppl.) {\bf 120} (1995) 57.

\item{\rm 2.} H.~Ichie, H.~Suganuma and H.~Toki, 
Phys.~Rev.~{\bf D52} (1995) 2944.

\item{\rm 3.} H.~Ichie, H.~Monden, H.~Suganuma and H.~Toki, 
Proc.~of {\it Nuclear Reaction Dynamics of Nucleon-Hadron Many Body 
Dynamics}, Osaka, Dec. 1995, in press.

\item{\rm 4.} H.~Suganuma, A.~Tanaka, S.~Sasaki \& O.~Miyamura, 
Nucl.~Phys.~{\bf B}(Suppl.){\bf 47} (1996) 302.

\item{\rm 5.} T.~Kajino, M.~Orito, T.~Yamamoto and H.~Suganuma, 
{\it Confinement '95}, 
(World Scientific, 1995) 263.

\end